\documentstyle[12pt]{article}

\title{Trace Analysis of Ancient Gold Objects Using Radiochemical Neutron 
Activation}
\author{Agata Olariu$^*$, Mioara Constantinescu, O. Constantinescu,\\
T. Badica, I. V. Popescu\\
National Institute of Physics and Nuclear Engineering\\
P.O.Box MG-6, 76900 Magurele, Bucharest, Romania\\
$^*$e-mail: agata@tandem.ifa.ro\\
C. Besliu\\
Faculty of Physics, Bucharest\\
Doina Leahu\\
National History Museum, Bucharest\\}
\begin{document}
\date{}
\maketitle

\begin{abstract}

Radiochemical neutron activation analysis  has been 
applied to investigate the microelements in gold samples with archaeological 
importance. Chemical separation has allowed the determination of traces of
Ir, Os, Sb, Zn, Co, Fe, Ni. Instrumental neutron activation analysis has been
used for the determination of Cu.

\end{abstract}
\newpage
\setcounter{page}{2}
\setlength{\topmargin}{-0.5cm}
\headheight 0cm
\setlength{\textwidth}{13cm}
\setlength{\textheight}{23.5cm}

\begin{center}
\large
{\bf Introduction}\\
\end{center}
\large
\noindent
The chemical composition 
of any archaeological item can be used
to develop characterisation and classification criteria $^1$.
In this work we have analysed a number of ancient
gold objects from different gold hoards from the National History Museum, 
Bucharest, to answer some questions related to the composition
of these objects. We present in Table 1 the list of gold samples of historical 
interest which have been studied in this work.
Investigations of Emeleus$^2$ and Caley$^3$ indicate that the non-destructive
instrumental neutron activation analysis (INAA)
is not suitable for the  determination of various 
impurities in gold. Indeed, because of the spectroscopic interference effects
 or the 
low sensitivity of INAA, many
impurities in gold matrix cannot be detected. Among the perturbative effects,
the Ag with high concentrations in gold samples is a serious problem. 
The ancient gold samples 
contain relatively high concentrations of Ag
$>1\%$, and after irradiation with thermal neutrons the gamma spectra of the 
samples,
after the
disintegration  of gold, are dominated by the $\gamma$
rays of the isotope 
$^{110m}$Ag (T$_{1/2}$=250 d). 
\newpage
We have concluded that for the investigation of ancient gold samples by 
NAA, the chemical separation of Ag is required, and
in this way the sensitivity of other minor elements in gold can be improved.
A post-irradiation
separation of the Ag from the gold sample, was chosen to avoid the 
 contamination from the reagents  used in the pre-irradiation separation 
procedure. RNAA has allowed
the determination of Ir, Os,
Sb, Zn, Co, Fe and Ni, in the ancient gold samples.  \\

\begin{center}
                        {\bf Experimental}\\
\end{center}
{\bf RNAA}. In order to determine some trace elements in ancient gold samples we have
applied a RNAA technique. After neutron irradiation, the silver was 
precipitated
as AgCl and centrifugated. The supernatant solution  includes the 
traces of elements in the gold samples.  \\
{\it Preparation of samples}: samples of $\approx$ 3mg were wrapped in 
pure Al foil. To diminish the self-shielding  effects of
neutron flux, the samples have been flattened$^4$. A standard of gold of 0.105
$\mu$g, 99.9\% was used.\\
{\it Irradiation}: samples and the standard have been irradiated for 30 h  
at a neutron flux of 
10$^{13}$ neutrons/cm$^{2}\cdot$ sec in the VVR-S reactor of the NPNE, 
Bucharest-Magurele.
\newpage
\noindent
{\it Preliminary measurements}: in the first days after the  irradiation, 
the gamma spectra  are dominated
by the photopeaks of $^{198}$Au and after the disintegration of this, the 
spectra are dominated by $^{110m}$Ag. This fact does not permit to determine
other elements than Au and Ag in the gold samples. 
After $\approx$30 - 40 d, the samples were counted to measure Au and Ag.
%the activity of $^{110m}$Ag was measured. 
The measurements have been performed with a Ge(Li) detector, 135 cm$^3$ and a
 Canberra 4096
channel analyser. The system gave a resolution of 2.4 keV at 1.33 MeV 
($^{60}$Co). Preliminary spectral analysis has demonstrated the high 
sensitivity of the classical INAA procedure for Au and Ag. \\
{\it Radiochemical separation procedure}: by applying the chemical separation 
technique of Silver, several 
microimpurities, also present in gold could also be detected. 
The diagram from Fig. 1 indicates our radiochemical separation procedure of 
Ag from  
irradiated gold samples. 
 The irradiated gold samples
were dissolved in  warm  "aqua regia". After that, AgNO$_{3}$ as carrier
and 4M HCl were added to this solution and silver was precipitated as AgCl. 
The
precipitate was centrifuged and washed 3 times with 2.0 ml distilled H$_2$O
till neutral pH. 
The aqueous phase from the washing of AgCl precipitate was added to the
solution from the silver chloride precipitation.
\newpage
\noindent
In order to obtain a high degree of purification,  the
precipitation as AgCl has been repeated 3 times. 
Finally, the supernatant
solution has been evaporated on a small plastic plate and counted. 
We determined the radiochemical yield for Au,
 by measuring the activity of Au in
samples before, and after the chemical separation; the main value of the 
radiochemical yield for Au was in the range of 
92\%-95\%. For the other elements we could not 
evaluate the radiochemical yield, because before the separation of Ag, these
elements could not be seen, they were covered by the activity of Ag.  
It is to
mention that the sample dissolution was difficult to access, 
therefore we have used the repeated dissolution with "aqua regia" and hot,
concentrated, HNO$_{3}$ and HCl. \\
{\it Measurements}: after the Ag precipitation, the gold phase
was measured. Gamma ray spectra were recorded for 20,000 s per sample. 
The concentrations of the microelements were determined by the measurement of
the activity of the following photopeaks: Ir (316 keV), Os (646 keV),
Sb (603 keV), Zn (1115.5 keV), Fe (1099 keV), Co (1332 keV), and
Ni (811 keV). The Nickel concentration was determined by the 811 keV peak
belonging to $^{58}$Co isotope, arising from the $^{58}$Ni(n,p)$^{58}$Co 
reaction. The instrumental errors of measurements are $<$20\% for Os 
and $<$7\% for Ir, Sb, Zn, Fe, Co and Ni. 
\newpage
\noindent
Under our experimental conditions,
the detection limits (i.e. the lowest concentration of the
element under consideration that produces a peak under which the
integrated area is equal to 3 times the square root of the
background number of counts) are as follows:
Ir: 2 - 10 ppb, Os: 1 - 4 ppm, Sb: 0.3 - 1 ppm, Zn: 1 - 6 ppm, 
Fe: 150 - 2000 ppm, Co: 0.05 - 0.10 ppm, Ni: 100 - 300 ppm.  \\
{\bf INAA}. Copper was not possible to be determined by the RNAA procedure 
and a special
separate measurement was done: the samples of gold with  Au and Cu pure 
99.9\% standards
have been irradiated at a
flux of 10$^{12}$neutrons/cm$^{2}\cdot$sec with a rabbit system from the same 
reactor, for 2 minutes. After a cooling time of 3$\cdots$4
minutes,
the samples were counted for 500 s, with a lead brick of a thickness of 2.5 cm 
interposed between the detector and the sample. The standard was counted after
30 min, in the same geometry as the sample. The radioactivity in 
the photopeak of 412 keV of $^{198}$Au is attenuated $\approx$300 times and 
this fact
has permitted to observe the photopeak of 1039 keV of $^{66}$Cu 
(T$_{1/2}$=5.1 min), which is attenuated $\approx$ only  5  times  by  the 
lead
 brick.
The instrumental errors of measurements are $<$1\% for Au and Ag, 
and $<$20\% for Cu. The limit of detection for Cu of this
procedure was about 0.2\%.\\

\begin{center}
                       {\bf Results and Discussions}
\end{center}
The experimental results concerning the elemental concentrations of
ancient gold objects belonging to the National History Museum, Bucharest
are presented in Table 2. The concentrations are expressed in \% for the Ag 
and
Cu contents, in ppb for Ir and in ppm for Os, Sb, Zn, Co, Fe and 
Ni. The instrumental errors of measurements are $<$20\% for Os and
also for Cu, and $<$7\% for the other elements.
The interpretation of the compositional scheme of gold items is complicated 
$^5$
by the fact that the gold objects were frequently remelted during the history,
but less for jewellery than for coins. Also the representativity of a sample 
of $\approx$1-3mg, is questionable.
 Gold can be very inhomogeneous if it has not
been melted.$^6$
%[ E. Pernicka ]. 
However based on several thousands of analyses 
of ancient gold objects using samples of about 1 mg, Hartmann was able to 
distinguish different types of gold compositions with geographically coherent
distribution$^7$.\\
For Au, Ag and Cu distribution in the gold samples, 
Fig. 2 represents Ag/Au and Cu/Au plots.
The samples A1, A2 and A3 from Moigrad, Neolithic present close concentrations
of Au, Ag and Cu with the samples from Moigrad 5th century hoard:
A25, A26, A27 samples. The objects of the two hoards from Moigrad have the same 
discovery place and  are dated by the historians very different, at a long 
distance 
(Table 1).  
\newpage
\noindent
From the Fig. 2 it seems these gold hoards have the same 
major element composition.
The sample of A24, the buckle has a different
concentration of Cu (C$_{Cu}$=6.89\%) in comparison with the other items
from Moigrad, with the concentration lying in the range 
$<1\%.$ The sample A24  is also
distinguished with its higher concentrations  of  Ir C$_{Ir}$=455 ppb,
Zn C$_{Zn}$=3400 ppm and Ni C$_{Ni}$=230 ppm.\\
In Fig. 2 we have also represented the ratio Ag/Au versus Cu/Au, for some 
items
from the reference 7., items which are analysed also in the present work: 
the 3 idols Moigrad Neolithic (samples A1, A2 and A3). Our results are in a 
relative good agreement with the results of reference 7. 
The samples from Arge\c s hoard, beads and rings present a wide range of
concentrations as concerning Cu, but also they can be marked on the graph 
like a relatively separate region.\\
In Fig. 3 Sb -  Zn correlation is shown for some samples. 
The samples from Arge\c s hoard form a relatively close 
cluster. For the beads $\bar C_{Zn}$=328$\pm$61 ppm, and 
$\bar C_{Sb}$=14$\pm$5 ppm\\
and for the rings $\bar C_{Zn}$=190$\pm$86 ppm, and 
$\bar C_{Sb}$=24$\pm$8 ppm\\ 
From the elements revealed in ancient gold samples, the elements
Fe, Co and Ni compose a relatively aleatory  background, traces of these
 elements
can be added or removed during the manufacturing or refining process. \\
One can observe the presence of Fe in the gold items from Turnu 
Magurele.\\
The importance of noble metals in the gold sample is well known$^{8,9}$. 
The noble metals remain in unchanged concentrations  during  the  processing 
or refining. In our samples, Ir and Os have been determined in very low 
concentrations. 
%In the reference$^9$ Ir have been also determined in low 
%concentrations for the native Romanian gold $^9$.\\
%C$_{Ir}$=410$\pm$220 ppm -Ro\c sia Montan\u a, Apuseni Mountains\\
%C$_{Ir}$=640$\pm$270 ppm -Pian Valley, ''\\
%C$_{Ir}$=44$\pm$44 ppm -Ruda Brad, ''\\
%By the low concentration of the iridium,
%it suggests the belonging of the analysed items to the autochthon gold.\\
From the analysed objects we could remark some items with higher 
concentrations of Ir: bracelet 
Some\c seni, sample A35: C$_{Ir}$=1320$\pm$90 ppb and bracelet with muff, 
sample A36: C$_{Ir}$=520$\pm$40 ppb.\\

In this report it was shown that RNAA  can be
used as a method for the 
study of elemental composition of ancient gold objects. By this method a 
number of elements like: Ir, Os, Sb, Zn, Co, Fe and Ni can be determined in 
the ancient gold which improves the knowledge of historians concerning these 
valuable objects.

\newpage
\begin{center}
                                {\bf References}
\end{center}
\noindent
%\begin{thebibliography}{}
1. Z. Goffer, 1980, Archaeological Chemistry, N.Y., J. Wiley \& Sons, 1980\\
2. V. M. Emeleus, Archaeometry, vol. {\bf 1} (1958) 6\\
3. E. R. Caley, Analysis of Ancient Metals, Oxford, Pergamon Press, 1964\\
4. T. Takeutchi, Radioisotopes {\bf 29} (1980) 119\\
5. F. Widemann, 1980, J. Radioanal. Chem. Vol. {\bf 55} (1980) 272\\
6. E. Pernicka, Nucl. Instr. and Methods {\bf B 14} (1986) 24\\
7. A. Hartmann, Prehistorische Goldfunde aus Europa II, Mann. Berlin, 1982\\
8. N. D. Meeks and M. S. Tite, Journal of Archaeological Science,
   {\bf 7} (1980) 267\\
9. V. Cojocaru, 3$^{rd}$ General Conf. of the Balkan Phys. Union, Sep. 1997,
Cluj, Romania, Abstracts, p.495\\
%\end{thebibliography}
\normalsize
\begin{table}
\caption{Table 1 List of ancient gold objects analysed in this work}
\bigskip
\begin{tabular}{ll}
\hline
Sample  & Ancient gold object, hoard, provenance, dating, inventory number\\
\hline
\hline
A1      & "Violin"- Idol, Moigrad, jud. S\u alaj, 3 rd millenium, B.C.,
inv.no. 54570\\
A2      & Masc. Idol Moigrad, 3 rd millenium, B.C., inv.no. 54571\\
A3      & Femin. Idol Moigrad, 3 rd millenium, B.C., inv.no. 54572\\
A4      & Bracelet, Lungoci, jud. Gala\c ti, inv.no. 89336\\
A5      & Disk Phalera, Ostrovul Mare, jud. Mehedinti, 15 th century B.C., 
inv.no. C456\\
A6      & Disk Phalera, Ostrovul Mare, inv.no. C 473\\
A10     & Ring "spiral", Arge\c s, 15 th - 14 th centuries B.C., inv.no. C 
429\\
B3-B9   & Beads, Arge\c s\\
B10-B19 & Rings "spiral", Arge\c s\\
A11     & Muff, Turnu M\u agurele, jud. Teleorman, 13 th -12 th centuries 
B.C., inv.no. 11039\\
A12     & Muff, Turnu M\u agurele, inv.no. 11042\\
A14     & Muff, Turnu M\u agurele, inv.no. 11040\\
A15     & Muff, Turnu M\u agurele, inv.no. 11043\\ 
A16     & Ring "spiral", Turnu M\u agurele, inv.no. 92037\\
A17     & Tubular Object, Turnu M\u agurele, inv.no. 11038\\
A18     & Bracelet, Dip\c sa, jud. Bistri\c ta-N\u as\u aud, 13 th-12 th
centuries B.C.,  P 23484\\
A19     & Diadem, Bune\c sti-Avere\c sti, jud. Vaslui, 3 rd century B.C.\\
A20     & Ingot, Feldioara, , jud. Bra\c sov, 379-380 A.D.\\

\end{tabular}
\end{table}

\begin{table}
\begin{tabular}{ll}
A21     & Diadem, Balaci, jud. Teleorman, 5 th century A.D.\\
A22     & Bracelet, Balaci, inv.no. 9133\\
A23     & Bracelet, Balaci, inv.no. 9134\\
A24     & Buckle Moigrad, 5 th century, jud. S\u alaj, inv.no. 54587\\
A25     & Bracelet Moigrad, 5 th century, inv.no. 54585, \\
A26     & Bracelet Moigrad, 5 th century, inv.no. 54584 \\
A27     & Bracelet Moigrad, 5 th century, inv.no. 54586\\
A28     & Ear, Moigrad, 5 th century, inv.no. 54588\\
A29     & Floral applique, Buz\u au, 5 th century, inv.no. 9014\\
A30     & Buckle Buz\u au, inv.no. 11060\\
A31     & Ear ring pendantiv, Buz\u au\\
A32     & Tubular object, Mangalia, 4 th century, inv.no. 8977\\
A33     & Mangalia, fragment necklace, inv.no. 8976\\
A34     & Piece, Mangalia, inv 8975\\
A35     & Bracelet, Some\c seni, Cluj, inv.no. 54252\\
A36     & Bracelet with Muff, inv.no. 11084\\
A38     & Bowl "with snails", Dobrogea, inv.no. 11091\\

\end{tabular}
\end{table}
\begin{table}
\caption{Concentration of elements in different ancient gold objects, by RNAA,
(Cu by INAA)}
\begin{tabular}{lccccccccc}
\hline
Sample & Ag     &      Cu       &       Ir      &      Os       &      Sb      
 &      Zn       &       Fe      &       Co      & Ni\\
        & \%    &      \%       &      ppb      &     ppm       &     ppm      
&   ppm          &    ppm         &      ppm     &  ppm \\       
\hline
\hline
A1&     4.98    &       0.523   &       24      &     $<$1    &       1.3    
 &      7.1     &       190     &       1.2     &128\\
A2&     6.01    &       0.404   &       8.8     &     $<$3    &       3.1    
 &      29      &       1420    &       0.8     &138\\
A3&     6.03    &       0.323   &     $<$10     &     $<$4    &       1.5    
 &      38      &      $<$2000         &       4.26    &$<$300\\
A4&     7.65    &       0.554   &       125     &     $<$8      &       5.2    
 &      9.8     &       -       &       0.67    &1050\\
A5&     21.8    &       0.303   &       -       &       -       &       -      
 &      42      &       -       &       0.26    &-\\
A6&     17.6    &       0.794   &       -       &       -       &       -      
 &      29      &       -       &       -       &-\\
A7&     21.6    &       0.351   &       11      &       23      &       24     
 &      40      &       760     &       -       &-\\
A8&     20      &       0.408   &       67      &     $<$20     &       $<$1   
 &      15      &       -       &       0.015   &-\\
A9&     11.6    &       0.315   &       3       &       -       &       0.94   
 &      14      &       -       &       0.93    &-\\
A10&    20      &       0.554   &       -       &       -       &       1.9    
 &      12      &       -       &       0.12    &-\\
B3&     24.7    &       1.03    &       -       &       -       &       12.3   
 &      430     &       -       &       -       &-\\
B4&     30.5    &       0.725   &       -       &       -       &       8.4    
 &      309     &       3980    &       -       &-\\
B5&     27.3    &       0.625   &       324     &       34      &       22     
 &      490     &       6025    &       -       &-\\
B6&     11.2    &       58.7    &       -       &       25      &       3.8    
 &      125     &       1890    &       -       &-\\
B7&     23.6    &       15.5    &       -       &       17      &       7.7    
 &      118     &      2280     &       -       &230\\
B8&     26.5    &       3.71    &       30      &       30      &       20     
 &      309     &       5374    &       -       &-\\
B9&     30.3    &       0.998   &       -       &       114     &       23     
 &      518     &       5220    &       -       &-\\
B10&    17.9    &       0.554   &       -       &       242     &       78     
 &      258     &       5111    &       -       &-\\
B11&    14.7    &       46.1    &       -       &       7       &       8      
 &      76      &       1290    &       -       &-\\
B12&    24.4    &       11.1    &       -       &       -       &       3      
 &      195     &       1104    &       -       &-\\
B13&    27.5    &       2.75    &       -       &       24      &       24     
 &      152     &       1846    &       -       &-\\
B14&    20.6    &       0.610   &       15      &       8.7     &       6.1    
 &      142     &       1364    &       -       &-\\
B15&    13.3    &       45.1    &       -       &       -       &       5.2    
 &      15      &       -       &       -       &-\\
B17&    18.5    &       38.4    &       -       &       -       &       32     
 &      762     &       -       &       -       &-\\
B18&    33.6    &       0.399   &       -       &       -       &       16.5   
 &      54      &       -       &       -       &-\\
B19&    30.0    &       0.289   &       -       &       -       &       46     
 &      53      &       -       &       -       &-\\
\end{tabular}
\end{table}

\begin{table}
\begin{tabular}{lccccccccc}
\hline
Sample & Ag     &      Cu       &       Ir      &      Os       &      Sb      
 &      Zn       &       Fe      &       Co      & Ni\\
\hline
\hline
A11&    51.6    &       2.14    &       15      &       -       &       -      
 &      11      &       -       &       0.94    &-\\
A12&    17.1    &       0.571   &       91      &       -       &       5.7    
 &      30      &       -       &       -       &-\\
A14&    11.1    &       2.49    &       198     &       -       &       7.5    
 &      12      &       -       &       -       &-\\
A15&    15.9    &       0.285   &       36      &       -       &       0.7    
 &      1.7     &       -       &       -       &-\\
A16&    21.2    &       2.63    &       -       &       -       &       -      
 &      -       &       -       &       -       &-\\
A17&    16.0    &       0.9340  &       -       &       -       &       -      
 &       2.56    &       -       &       -       &-\\
A18&    10.7    &       0.337   &       98      &       -       &       -      
 &      62      &       -       &       0.98    &-\\
A19&    0.7702  &       0.6020  &       20.3    &       -       &       -      
 &       0       &       -       &       -       &-\\
A20&    0.74    &        -      &       140     &       -       &       -      
 &      0       &       -       &       -       &-\\
A21&    12.7    &       1.84    &       145     &       -       &       -      
 &      30      &       -       &       1.8     &-\\
A22&    8.09    &       0.389   &       -       &       -       &       -      
 &      23      &       -       &       0.94    &-\\
A23&    10.8    &       -       &       35      &       -       &       -      
 &      44      &       -       &       -       &-\\
A24&    8.32    &       6.89    &       455     &       11      &       -      
 &      3400    &       -       &       0.078   &230\\
A25&    9.44    &       0.4147  &       -       &       -       &       1      
 &      16      &       -       &       0.676   &-\\
A26&    6.42    &       0.522   &       -       &       -       &       7.2    
 &      14      &       -       &       1.1     &-\\
A27&    7.21    &       0.729   &       -       &       -       &       2.2    
 &      15      &       -       &       0.738   &-\\
A28&    5.62    &       0.314   &       149     &       -       &       1      
 &      54      &       -       &       4.47    &420\\
A29&    11.5    &       4.000   &       72      &       -       &       3.6    
 &      615     &       -       &       0.89    &-\\
A30&    3.05    &       0.486   &       250     &       -       &       2.4    
 &      230     &       -       &       -       &-\\
A31&    8.24    &       5.14    &       44      &       -       &       0.18   
 &      19      &       -       &       -       &-\\
A32&    10.6    &       4.06    &       420     &       -       &       2.2    
 &      27.3    &       -       &       2.4     &-\\
A33&    2.92    &       6.26    &       587     &       -       &       2.1    
 &      24.7    &       -       &       1.5     &250\\
A34&    3.71    &       0.242   &       488     &       -       &       8.6    
 &      35      &       -       &       2.3     &-\\
A35&    4.74    &       0.646   &       1320    &       -       &       11.7   
 &      61      &       -       &       2.6     &-\\
A36&    5.75    &       3.13    &       520     &       -       &       12.6   
 &      45      &       -       &       1.1     &-\\
A38&    13.44   &       8.62    &       403     &       -       &       44     
 &      70      &       -       &       0.9     &-\\
\hline
\end{tabular} 
{\scriptsize
 
- = under the detection limit\\}
\end{table}
\vspace*{9cm}
\large
FIGURE CAPTIONS

Fig. 1 Diagram of the radiochemical separation procedure\\

Fig. 2 Ratio Ag/Au versus the ratio Cu/Au for different ancient gold items, by 
INAA\\

Fig. 3 Concentration of Sb versus the concentration of Zn for different 
ancient
       gold objects, by RNAA\\

\newpage
TABLE CAPTIONS\\

Table 1 List of analysed samples.\\

Table 2 Concentration of elements in different ancient gold objects, by RNAA.
 The concentrations are expressed in \% (Ag and Cu), in ppb (Ir) and in ppm
(Os, Sb, Zn, Fe, Co and Ni)\\

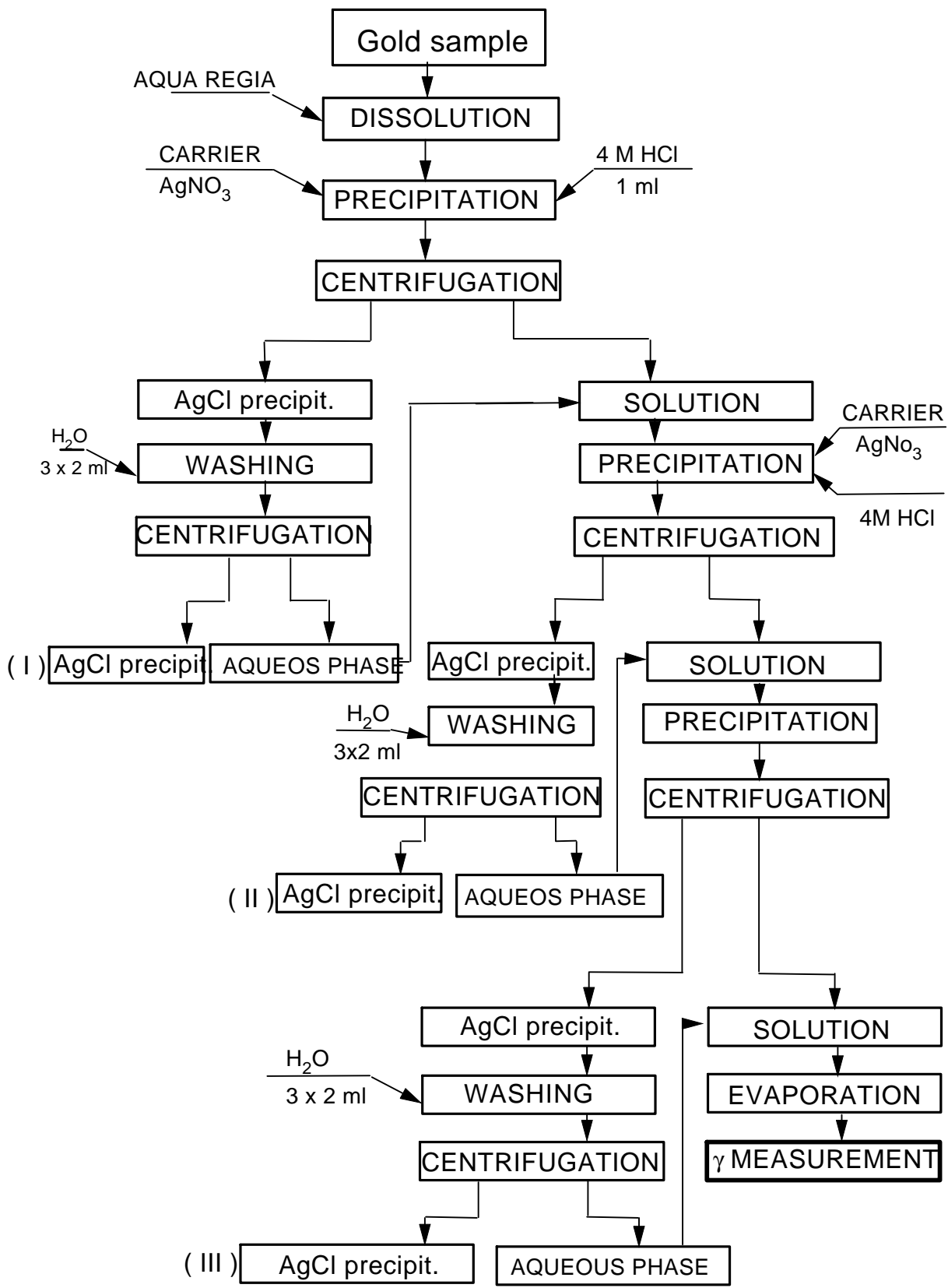
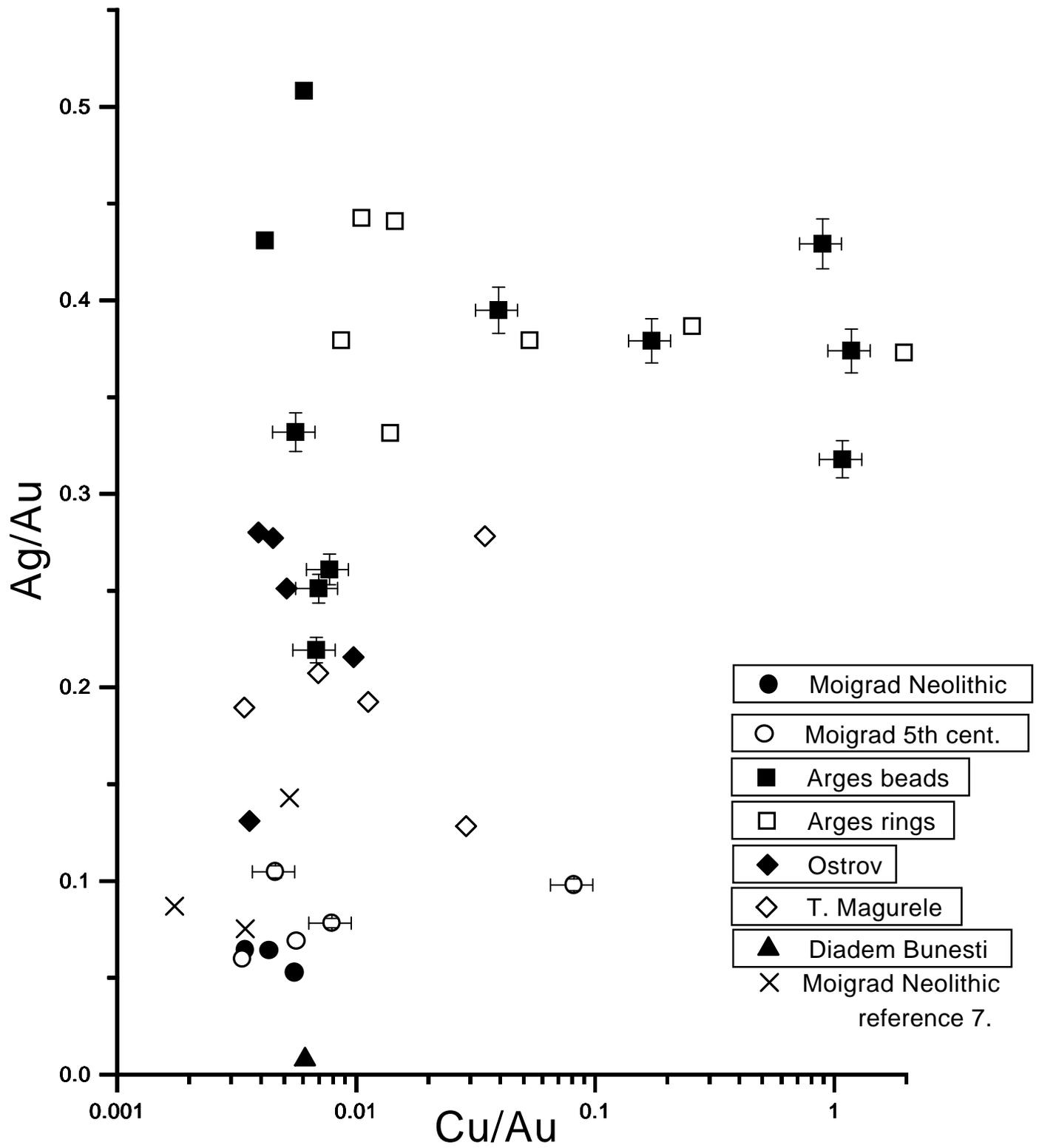
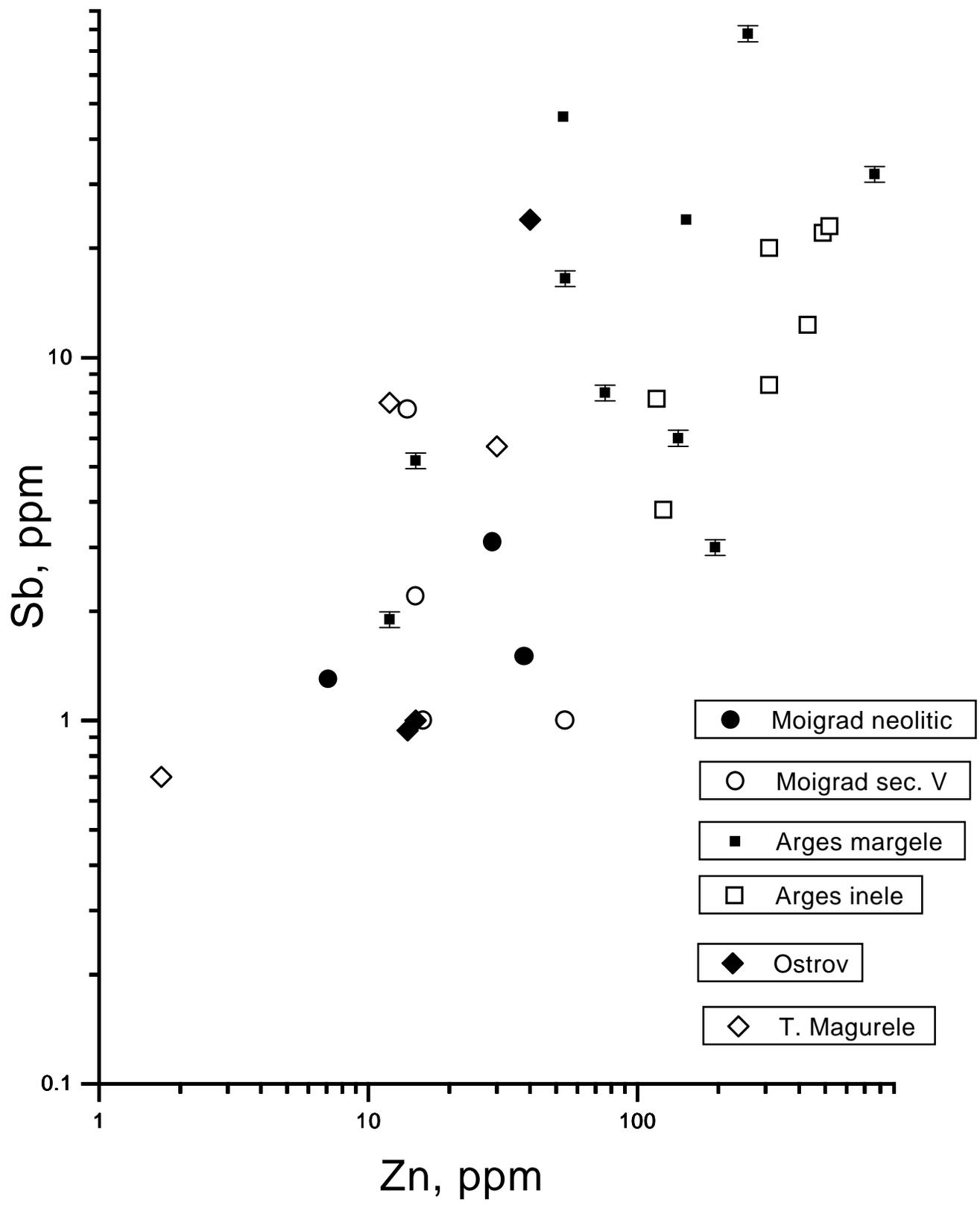
\end{document}